\documentclass[aps,pra,showpacs,twocolumn,superscriptaddress,floatfix]{revtex4-1}
\usepackage{graphicx,color}
\usepackage{bm}
\begin{document}
\title{Ground and excited state properties of the polar and paramagnetic RbSr molecule: a comparative study.}
\author{Piotr S. \.Zuchowski}
\email{pzuch@fizyka.umk.pl}%
\affiliation{Instytut Fizyki, Uniwersytet Miko\l aja Kopernika, ul. Grudziadzka 5/7, 87-100 Torun, Poland}
\author{R. Gu\'erout}
\email{romain.guerout@spectro.jussieu.fr}%
\affiliation{Laboratoire Kastler-Brossel, CNRS, ENS, UPMC, Case 74, F-75252 Paris, France}
\author{O. Dulieu}
\email{olivier.dulieu@u-psud.fr}%
\affiliation{Laboratoire Aim\'e Cotton, CNRS/ Univ. Paris-Sud/ ENS-Cachan, B\^at. 505, Campus d'Orsay, 91405 Orsay Cedex, France.}

\date{\today}

\begin{abstract}
This paper deals with the electronic structure of RbSr, a molecule possessing both a permanent magnetic and electric dipole moment in its own frame allowing its manipulation with external fields. Two complementary \textit{ab-initio} approaches are used for the ground and lowest excited states: first, an approach relying on optimized effective core potentials with core polarization potentials based on  a full configuration interaction involving three valence electrons, and second, an approach using a small-size effective core potential with 19  correlated electrons in the framework of coupled-cluster theory. We have found excellent agreement between these two approaches for the ground state properties including the permanent dipole moment. We have focused on studies of excited states correlated to the two lowest asymptotes Rb($5p\, ^2\!P$)+Sr($5s^2\,^1\!S$) and Rb($5s\,^2\!S$)+Sr($5s5p\,^3\!P$) relevant for ongoing experiments on quantum degenerate gases. We present also the Hund c) case potential curves obtained using atomic spin-orbit constants. These potential curves are an excellent starting point for experimental studies of molecular structure of RbSr using high-resolution spectroscopy.
\end{abstract}

\pacs{34.20.-b, 34.50.Cx, 37.10.Pq} 

\maketitle
\section{Introduction}
The detailed investigation of the properties of  quantum degenerate gases of ultracold species  (\textit{i.e.} with kinetic energy $E_k \equiv k_BT$ equivalent to a temperature $T \ll 1$~millikelvin) is among the most important goals of modern atomic, molecular, optical and statistical physics. A unique feature of ultracold quantum gases is the tunability of the interaction strength between the particles with the external  fields:  by employing the Feshbach resonances \cite{Chin:2010} it is possible to change the  scattering length in  broad range of values. By comparison with atoms, the rich internal structure of polar molecules (i.e. possessing a permanent electric dipole moment) and their mutual strong anisotropic interactions can offer to this field novel opportunities for precision measurements and for quantum control using electromagnetic fields \cite{Carr:2009,Dulieu:2009}.
Ultracold molecules trapped in periodic optical lattices have been proposed as qubits for prototypes of quantum computers \cite{DeMille:2002}, or as quantum simulators for studies of many-body phenomena such as phase transitions, strongly correlated systems or many-body physics in reduced dimensions \cite{Micheli:2006,Bloch:2008}.   
In 2008 two groups have reported the formation of ultracold gases of  polar LiCs and KRb molecules in ultracold temperatures \cite{Deiglmayr:2008,Ni:KRb:2008}:  ultracold LiCs molecules have been obtained by photoassociation of pairs of ultracold  Li and Cs atoms and spontaneous decay of excited LiCs$^*$ molecule down to the electronic ground state, while  ultracold KRb molecules have been created through magnetoassociation of ultracold K and Rb atoms into weakly bound  levels of the molecular ground state, followed by stimulated Raman adiabatic passage (STIRAP) toward the lowest rovibrational level \cite{Ospelkaus:2008,Ni:2008}. There is also a number of other experiments aiming at creating ultracold heteronuclear diatomic alkali-metal molecules in their ground state like RbCs \cite{Debatin:2011,Takekoshi:2012} and NaK \cite{Wu:2012:NaK}, since  in contrast with KRb \cite{Ospelkaus:2010} they are stable with respect to the chemical reactions of atom exchange and trimer formation \cite{Zuchowski:2010a}.

Heteronuclear diatomic alkali-metal molecules in their ground state $X^1\Sigma^+$, however, are not easy to manipulate with external fields: their very weak magnetic moment originates only from nuclear spin, and they do not exhibit a linear Stark effect in the rovibrational ground state. A very interesting class of quantum simulators has been proposed by Micheli {\em et al.} \cite{Micheli:2006}  employing molecules with both an electric and magnetic permanent dipole moment in their own frame. Such molecules  reveal fascinating potential for high-precision measurements (for example the YbF molecule is being used in the determination of bounds for the electric dipole moment of the electron \cite{Hudson:2002}) or for sensitive imaging of low-frequency electromagnetic fields \cite{Alyabyshev:2012}.  In the rest of the paper we will qualify in short such species as paramagnetic and polar molecules.

One of the possible candidates for paramagnetic and polar molecules are diatomic molecules formed by association of laser-coolable atoms with different atomic spin quantum numbers, such as pairs of alkali-metal atoms and alkaline-earths atoms  \cite{Guerout:2010,Zuchowski:2010:RbSr}. One of the most promising candidates for such system is the RbSr molecule. Besides its magnetic doublet $X^2\Sigma^+$ electronic ground state, it exhibits a permanent electric dipole moment of about 1.4-1.5~Debye \cite{Zuchowski:2010:RbSr,Guerout:2010}. The laser cooling, trapping and manipulation of Rb atoms have been well-established at the very beginning of the  ultracold matter studies \cite{Anderson:1995}. At present the strontium atom is one of the most popular atomic species in ultracold physics \cite{michelson2005,nagel2005}:
for example, the studies of  Bose-Einstein condensation of Sr atoms and Bose-Fermi mixtures (of different Sr isotopes)  has recently been reported \cite{Stellmer:2009,Stellmer:2012,Stellmer:2013,Stellmer:2010,Tey:2010}. Moreover, the Innsbruck group has developed the STIRAP scheme to produce weakly bound Sr$_2$ molecules in ground electronic state. It is finally worth mentioning that Sr$_2$ molecules have also been produced by spontaneous decay from excited Sr($^1S$)--Sr($^3P_1$) molecular state \cite{Reinaudi:2012}. More recently a quantum degenerate gas of rubidium atoms coexisting with strontium has been produced \cite{Pasquiou:2013}. Another motivation which makes the studies of RbSr system particularly interesting is the magnetic tunability of scattering length  due to presence of subtle mechanisms which can produce the Feshbach resonances \cite{Zuchowski:2010:RbSr}. That might  allow the experimentalist to modify the scattering length in the ultracold mixture of Rb and Sr and control the behaviour of quantum gas of such atoms. It is worth mentioning that several other similar species are subject to intense ongoing research, like YbLi \cite{Okano:2010, Hara:2011, Hansen:2011,Zhang:2010,Gopakumar:2010} and YbRb \cite{Sorensen:2009,Nemitz:2009,Borkowski:2013}.

Manipulation of the quantum states of diatomic molecules with laser light requires  the knowledge of appropriate transition energies, and thus of the potential energy curves (PECs) supporting the relevant energy levels and the corresponding transition dipole moments (TDMs). Surprisingly enough, still only little is known about the structure of molecules containing alkali-metal atoms with  group II atoms. The electronic structure of Ba neutral compounds (BaLi, BaNa, BaK) has been explored some time ago by Allouche and coworkers \cite{Allouche:1994:BaLi,Boutassetta:1994,Boutassetta:1995}. Other studies concern CaLi \cite{Allouche:1994:CaLi,Russon:1998,Ivanova:2011}, LiBe \cite{Pak:1991,Marino:1992}. More recently the electronic structure of the related molecular ions containing one alkali atom and Ca$^+$ \cite{Hall:2013:a}, Sr$^+$ \cite{Aymar:2011} or Ba$^+$ \cite{Knecht:2008,Knecht:2010,Krych:2011,Hall:2013:b} with various high-level approaches have been published in relation with experiments aiming at creating cold molecular ions in merged cold ion and cold atom traps.

In this paper we present the studies of interactions of Rb  and Sr atoms in ground and excited states.
We have recently examined this system  in its $X^2\Sigma^+$ ground state with two entirely different approaches  \cite{Zuchowski:2010:RbSr,Guerout:2010}: one relies on the representation of RbSr as a three-valence-electron molecule in the field of relativistic polarizable large effective core potentials (ECPs) through a full configuration interaction (FCI) calculation, while the other treats explicitly 19 electrons in the field of a relativistic small core ECP via  the coupled cluster (CC) theory. Here, we use these  approaches to revisit  and to extend the study of the electronic structure of the RbSr system. The methods are described in Section \ref{methodssec}. The ground state properties of RbSr are carefully revisited in Section \ref{sec:ground} including the potential curve, the permanent dipole moment, and the static dipole polarizability. We have calculated the PECs and the transition dipole moments between the $X^2\Sigma^+$ (Rb($5s\,^2S$)+Sr($5s^2\,^1S$)) ground state and the excited $^2\Sigma^+$ and $^2\Pi^+$ states correlated to the two lowest asymptotes Rb($5p\,^2P$)+Sr($5s^2\,^1S$) and Rb($5s\,^2S$)+Sr($5s5p\,^3P$) relevant for the ongoing experiments (Section \ref{sec:excited}). We have also investigated the spin-orbit (SO) coupling of these states within the framework of an atomic model involving experimental atomic SO splittings. The results of this paper are of key importance in experimental investigations of the spectroscopy and dynamics of the RbSr diatom. These data could be adjusted to spectroscopic data in order to provide the essential information for designing optical routes for the formation of ultracold ground-state RbSr molecules in their lowest internal level, thus opening the way toward a degenerate quantum gas of molecules with both magnetic and electric dipole moments. 
In the rest of the paper, atomic units for distances (1~a.u. = $a_0$=0.0529177~nm), energies (1~a.u.=$2R_{\infty}$=219474.63137~cm$^{-1}$), and dipole moment (1~a.u.=2.541 580 59 Debye) will be used, except otherwise stated.
\section{ Methods }
\label{methodssec}
The first method is identical to the one used in our previous works for RbSr$^+$ \cite{Aymar:2011} and for the RbSr ground state \cite{Guerout:2010}. It is based on the representation of the Rb$^+$ and Sr$^{2+}$ ionic cores by relativistic effective core potential (ECP) complemented with core polarization potential (CPP) simulating core-valence correlation along the lines developed by M\"uller and Meyer \cite{Muller:CPP1:1984,Muller:CPP2:1984} and Foucrault \textit{et al.} \cite{Foucrault:1992}.These effective potentials involve semi-empirical parameters (reported in Ref.\cite{Guerout:2010}) which are chosen to reproduce the energies of the lowest $s$, $p$ and $d$ levels of the Rb and Sr$^+$ one-valence-electron systems. A full configuration interaction method (FCI) involving three valence electrons is performed in the framework of the CIPSI method (Configuration Interaction by Perturbation of a multiconfiguration wave function Selected Iteratively) developed at Paul Sabatier University in Toulouse (France). From now on we will refer to this method as FCI/ECP+CPP. Previous works on alkali dimers (see for instance Refs.\cite{Aymar:2005,Aymar:2007,Allouche:2012}) have demonstrated that this approach yields results for equilibrium distances ($R_e$) and potential well depths ($D_e$) for ground and excited states in good agreement with those obtained from experiments: for example, the discrepancy on $D_e$ for the $^1\Sigma^+$ ground states of alkali dimers is typically much less than 100cm$^{-1}$, often (eg. for KRb) less than 20 cm$^{-1}$. Values for permanent electric dipole moments (PEDMs) of their $^1\Sigma^+$ ground state \cite{Aymar:2005} match those measured in recent ultracold molecule experiments (\textit{e.g.} for KRb \cite{Ni:2008}, LiCs \cite{Deiglmayr:2010}). The TDMs functions are also found in close agreement with other theoretical values \cite{Aymar:2007,Aymar:2009,Kruzins:2010}. The (well-known) main advantage of the FCI/ECP+CPP method is its versatility and robustness: several low-lying  excited states can be easily calculated regardless their total spin, in contrast to single-reference quantum chemistry methods we employ in this paper (see below). Among disadvantages is its rapid increase of computational cost with increased basis set size. Just as in Ref.\cite{Guerout:2010} the basis set used in these calculations was limited to $s$, $p$ and $d$ Gaussian-type basis functions which translates into a number of configurations of about $10^5$. In the present case, the lack of $f$ orbitals mostly affects the evaluation of the dispersion interaction \textit{i.e.} its dependency in $R^{-6}$ is well reproduced, but its magnitude may not be correct.
 The basis-set superposition error (BSSE) has not been introduced, as we have checked that it remains small (less than 1~cm$^{-1}$) for the three valence electrons, while it is hard to estimate for the core electrons which are not explicitly taken in account.

The second method involves the calculations with fully relativistic small-core ECP (ECP28MDF) obtained by Lim {\em et al.} \cite{Lim:2006,Lim:2005}, such that all $4s$, $4p$ and $5s$ electrons (19) of Rb and Sr are correlated on both atoms.The PECs for the $X^2\Sigma^+$ ground state and for the lowest quartet $\Sigma^+$ and $\Pi$ states are determined within the open-shell spin-restricted coupled-cluster (RCC) theory \cite{Hampel:93} with single, double and triple excitations (RCCSD(T)) as in Ref.\onlinecite{Zuchowski:2010:RbSr} implemented in the {\sc Molpro 2012} package \cite{molpro:2012_brief}. In comparison to this work, we performed the calculations with significantly improved  basis set in order to estimate the error attributed to the basis set incompleteness. We used the original uncontracted basis sets of Lim {\em et al.} \cite{Lim:2006,Lim:2005} to which we have added $d$, $f$ and $g$ Gaussian-type basis functions to improve core-valence correlation between the $4s$ and $4p$ shells with the $5s$ one. We also added a series of diffuse $spdfg$ basis functions to better describe the dispersion interaction. We have further added $3s3p3d2f2g$ bonding functions.  We denote this new basis set with its maximum angular momentum $l_{\rm max}=4$. We have further extended this set to build a new one (denoted with $l_{\rm max}=5$) including one more large-exponent $g$ function for core-valence correlation  and one diffuse $h$ function. The latter has been used only for ground state calculations. Both basis sets are attached to this paper in the Supplementary material.

The doublet excited states have been obtained with the spin-restricted version of the open-shell equation-of-motion coupled cluster method limited to singly- and doubly- excited configurations(EOM-CCSD) \cite{Stanton:1993,Krylov:2008} implemented in the CFOUR package \cite{CFOUR_brief}. This approach allows for calculating excitation energies from the electronic ground state to the excited state  of any spatial symmetry, but it is unable to calculate the spin-flip transitions (and thus those involving the quartet states). The excitation energies of doublet states obtained with EOM-CCSD method has been then added to the ground state potential energy curve, while the lowest $1^4\Sigma^+$ and $1^4\Pi$ PECs have been shifted in order to smoothly match the Rb($5s\,^2S$)+Sr($5s5p\,^3P$) asymptotes calculated with EOM-CCSD. The basis functions has been restricted to the $l_{\rm max}=4$ set with removed bonding functions.

\section{The ground state properties of R\MakeLowercase{b}S\MakeLowercase{r}}
\label{sec:ground}
\begin{figure}
\caption{The RbSr ground state potential energy curves obtained with RCCSD (dashed red line), RCCSD(T) (full red line), and FCI/ECP+CPP methods (full blue line).}.
\includegraphics[width=\linewidth]{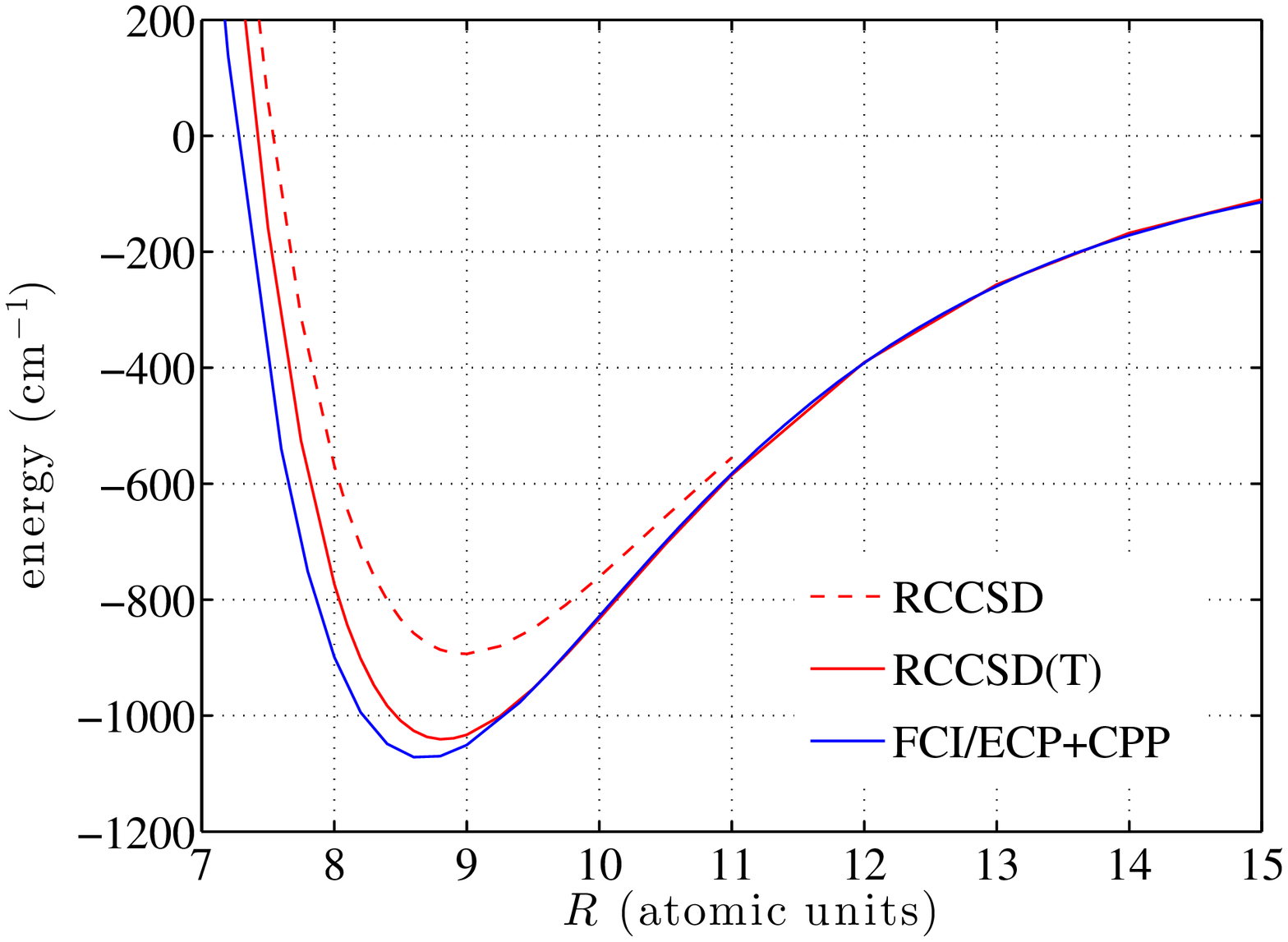}
\label{groundst}
\end{figure}

We present the results of our electronic-structure calculations for the ground state RbSr dimer using the RCCSD(T) method with both basis sets above in order to investigate the discrepancies between spectroscopic parameters obtained by Gu\'erout {\em et al.} \cite{Guerout:2010} and  \.Zuchowski {\em et al.} \cite{Zuchowski:2010:RbSr}, and to estimate the error bars due to the basis set truncation in CC calculations. Results are shown in Figure \ref{groundst}, while the essential spectroscopic parameters are gathered in Table \ref{gs_compare}.
%

The ground state potential calculated with the FCI/ECP+CPP method \cite{Guerout:2010} is $D_e=1073.3$~cm$^{-1}$ deep with an equilibrium distance $R_e=8.69a_0$ while in Ref. \onlinecite{Zuchowski:2010:RbSr} the depth of RbSr potential was found equal to 1000~cm$^{-1}$ at $R_e=8.86a_0$. With the new basis sets we have found that $D_e$ is approximately 3-4\% larger: the RCCSD(T) value is increased to $D_e=1034.4$~cm$^{-1}$ with the $l_{\rm max}=4$ basis set and to $D_e=1040.5$~cm$^{-1}$ with the $l_{\rm max}=5$ one. The difference in  $D_e$ in these basis sets is most likely related to a saturation of the dispersion energy in the calculations involving Gaussian functions. Based on well-know behavior of the correlation energy as a function of maximum angular momentum in the basis set \cite{Helgaker:1997:CBS} we can deduce the complete basis limit expected for RCCSD(T) method yielding a total interaction energy 1047.9~cm$^{-1}$. 
In fact, it is reasonable to treat the difference between extrapolated result and the interaction energy calculated using the $l_{\rm max}=5$ basis set as the uncertainty of the calculation. It is still quite hard to estimate the error beyond the RCCSD(T) calculation and to this end we will compare how analogous  methodology performs for the Sr$_2$ and 
singlet Rb$_2$ molecules. Skomorowski {\em et al.} have shown \cite{Skomorowski:2012:pra}, that the CCSD(T) dissociation energy of the Sr$_2$ dimer (1124~cm$^{-1}$), calculated with the same core potential and similar basis set, is slightly larger (by 3.8\%) than the experimental dissociation energy (1082 cm$^{-1}$).  The well depth of the Rb$_2$ ground state CCSD(T) PEC obtained with the ECP and basis set used in present study 
underestimates the experimental value (3836cm$^{-1}$) by 6\%. Thus, taking a 5\% uncertainty (52 cm$^{-1}$) on our basis set potential is certainly a conservative estimate. For completeness, we have also calculated the RbSr ground state PEC with spin-unrestricted coupled-cluster (UCC) approach: the potential depth for UCCSD(T) is no more than 20~cm$^{-1}$ larger than in the restricted case, which is within the estimated error bound. 

The result from the FCI/ECP+CPP calculation falls within such error bound, as the well depth is only 33~cm$^{-1}$ deeper than the RCCSD(T) value. The agreement between harmonic constants $\omega_e$ is also very good:  38.98 cm$^{-1}$ with the FCI/ECP+CPP approach and 38.09~cm$^{-1}$ with the RCCSD(T) calculations and the $l_{\rm max}=5$ basis set. Note, however, that $R_e$ is smaller by about 0.1~$a_0$ in the FCI/ECP+CPP approach than in the RCCSD(T).  This is probably due to the short-range repulsion between the Rb$^+$ and the Sr$^{2+}$ cores which is not automatically included the FCI/ECP+CPP approach invloving large cores. This contribution can be  represented by an exponential expression \cite{Pavolini:1989,Jeung:1997} fitted on the Hartree-Fock energy of the RbSr$^{3+}$ system.”

In fact both curves differ by merely one bound state and further two-color photoassociation spectroscopy for several isotopic mixtures of RbSr should provide the exact number of bound states supported by RbSr potential. The potential energy curves reported in this paper should be an excellent starting point for refinement using the experimental data. 
%

\begin{figure}
\caption{Permanent electric dipole moment of RbSr ground state calculated with finite-field method through the RCCSD(T) (full blue line) and the FCI/ECP+CPP approach \cite{Guerout:2010} (full red line).}
\includegraphics[width=\linewidth]{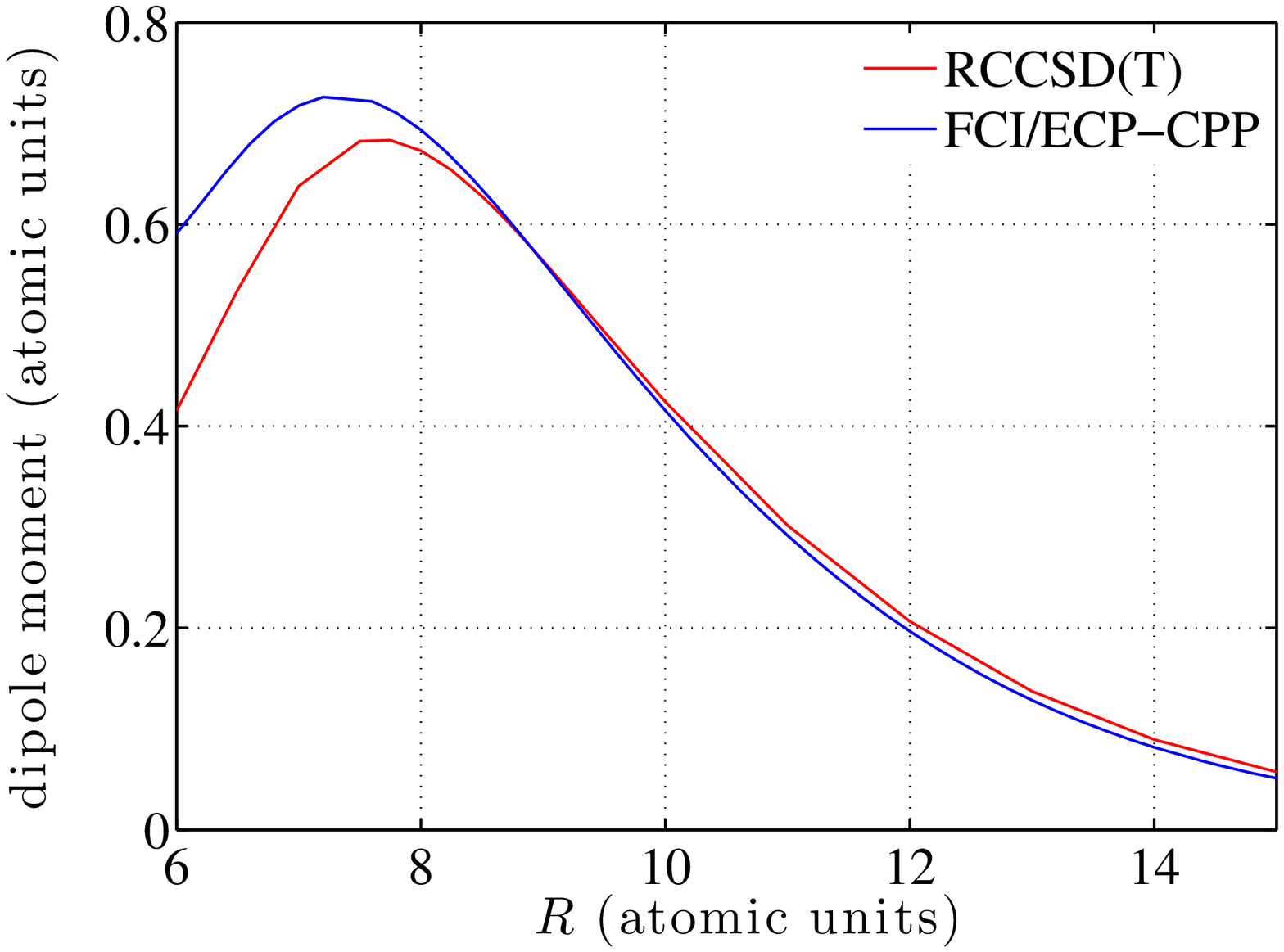}
\label{dipole_moment_plot}
\end{figure}

Other properties of the RbSr ground-state reveal the present quality of the electronic wave function when results are compared between two methods. Figure \ref{dipole_moment_plot} displays the ground state PEDM functions computed within the finite field approach, as obtained with the FCI/ECP+CPP method \cite{Guerout:2010}, and the new RCCSD(T) computation with extended basis set.
Both approaches yield very similar variation and magnitude, and thus very similar electronic wave functions. At the equilibrium distance the PEDMs are almost identical (1.54~D), and they become slightly different only at short internuclear distances. Note, that with previously reported calculations \cite{Zuchowski:2010:RbSr} - with the RCCSD(T) method employing smaller basis set - the value of the dipole moment was found to be 1.36~D.

\begin{figure}
\caption{Isotropic static dipole polarizability and the corresponding anisotropy of the RbSr ground state calculated with finite-field RCCSD(T) method.}
\includegraphics[width=\linewidth]{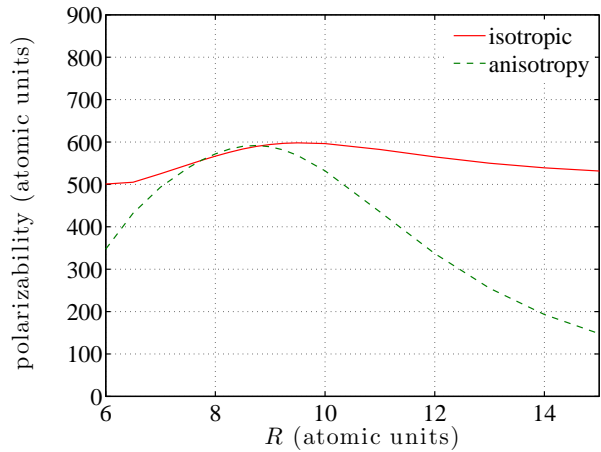}
\label{polarizability_gs}
\end{figure}

The similar finite-field approach  allows for calculating the static dipole polarizability of the RbSr ground state as the second derivative of the RCCSD(T) energy with respect to the amplitude of an external electric field.  We display in Fig. \ref{polarizability_gs} the $R$-dependent isotropic polarizability $\alpha_0$ and the corresponding anisotropy $\Delta\alpha$ which are related to the cartesian components according to the well-known formula 
\begin{equation}
\alpha_0=\frac{1}{3}(2\alpha_{xx}+\alpha_{zz}) \;\;\;  \Delta\alpha=\alpha_{zz}-\alpha_{xx}
\end{equation}
For the equilibrium distance the anisotropy of polarizability of RbSr molecule and the averaged polarizability are almost equal. The anisotropic polarizability peaks near $R_e$ while $\alpha_0$ has its maximum for 9.6 $a_0$. The $ \Delta\alpha$ at equilibrium distance is very large and comparable to the largest anisotropies reported for alkali-metal dimers \cite{Deiglmayr:2008,Zuchowski:2013}. With the large dipole moment and the large anisotropy of polarizability of RbSr, the RbSr molecule can be considered as a good candidate for manipulation with  intense off-resonant laser light \cite{Friedrich:1995,Tomza:2013,GonzalezFerez:2012}.


%

\begin{table}
\caption{ Equilibrium distance $R_e$ (in $a_0$) and potential depth $D_e$ (in cm$^{-1}$) of the RbSr ground state obtained by the various approaches discussed in the text.}
\begin{ruledtabular}
\begin{tabular}{crr}
 method &   $R_e (a_0)$  &  $D_e$ (cm$^{-1}$)   \\ \hline
  FCI/ECP+CPP, Ref. \onlinecite{Guerout:2010}           &   8.69          &  1073.3        \\    
RCCSD(T), Ref.  \onlinecite{Zuchowski:2010:RbSr}    & 8.86  &  999.6       \\ \hline
     \multicolumn{3}{c}{$l_{\rm max}=4$ basis set} \\ \hline
RCCSD      &   8.99  &  885.6       \\
RCCSD(T)      & 8.83  &  1034.4         \\
UCCSD     & 8.99  & 898.5          \\
UCCSD(T)    &  8.81 & 1052.5        \\ \hline
     \multicolumn{3}{c}{$l_{\rm max}=5$ basis set} \\ \hline
RCCSD    &  8.98   &   893.6         \\
RCCSD(T)    &  8.82 &  1040.5      \\
UCCSD        &   8.97   &  896.7          \\
UCCSD(T)    & 8.80 &  1059.1        \\
\end{tabular}
\end{ruledtabular}
\label{gs_compare}
\end{table}

\section{Excited states of the R\MakeLowercase{b}S\MakeLowercase{r} molecule}
\label{sec:excited}
\begin{figure}
\caption{ The diagram of experimental excited energy levels of Rb and Sr atoms featuring the corresponding dissociation limits (adding a ground state Sr atom on the left column, and a ground state Rb atom on the right column) of the molecular Hund's (a) case states of the RbSr molecule. The Rb $^2D_{3/2,5/2}$ energies of Rb are identical within the resolution of the plot. The origin of energies corresponds to infinitely separated ground state Sr and Rb atoms. }
\includegraphics[width=\linewidth]{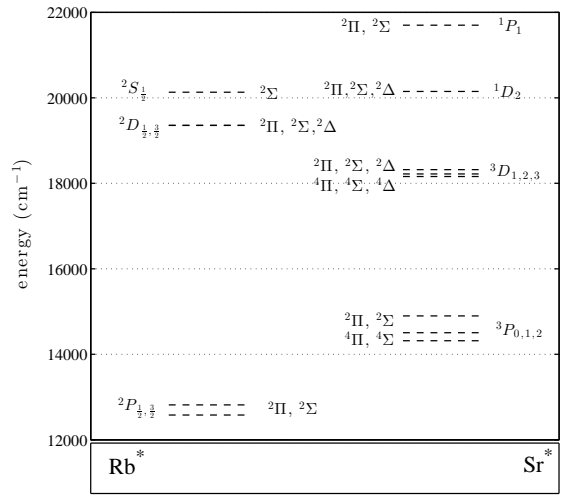}
\label{map_of_states}
\end{figure}

Figure \ref{map_of_states} shows the diagram of the  excited energy levels of Rb and Sr and lists the related Hund's (a) case states of RbSr. From the experimental point of view the most interesting excited states  are those correlating with the lowest asymptotes Rb($^2S$)+Sr($^3P_{0,1,2}$) and Rb($^2P$)+Sr($^1S$). In particular the  forbidden transition $^1S \rightarrow ^3P_1$ in Sr atom is very appealing for photoassociation experiments and optical manipulation, due its narrow width. This intercombination line has also been used recently for creation of ground state Sr$_2$ molecules \cite{Reinaudi:2012,Stellmer:2012}, and for optical tuning of the Sr scattering length \cite{Theis:2004,Ciurylo:2005,Nicholson:2011}. We  have also found that the states  correlated to the Rb($^2S$)+Sr($^3P$) asymptote might 
interact with higher excited states, thus we have also explored few of them -- namely the states which correlated to Rb($^2S$)+Sr($^3D$) and Rb($^2D$)+Sr($^1S$) asymptotes which are separated only by about 1000~cm$^{-1}$. Note that the Rb($6s\,^2S$) and the Sr($5s4d\,^1D)$ levels are very close to each other (20132.5~cm$^{-1}$ and 20149.7~cm$^{-1}$)
so that the {\em ab-initio} calculations are very difficult to perform,  regarding especially the proper order of asymptotic molecular states. The present approaches have been however successful in this matter.

\begin{figure*}
\caption{Hund's case (a) potential energy curves of the excited RbSr molecule obtained with EOM-CCSD method (left panel) and FCI/ECP+CPP method (right panel).}
\includegraphics[width=1.1\linewidth]{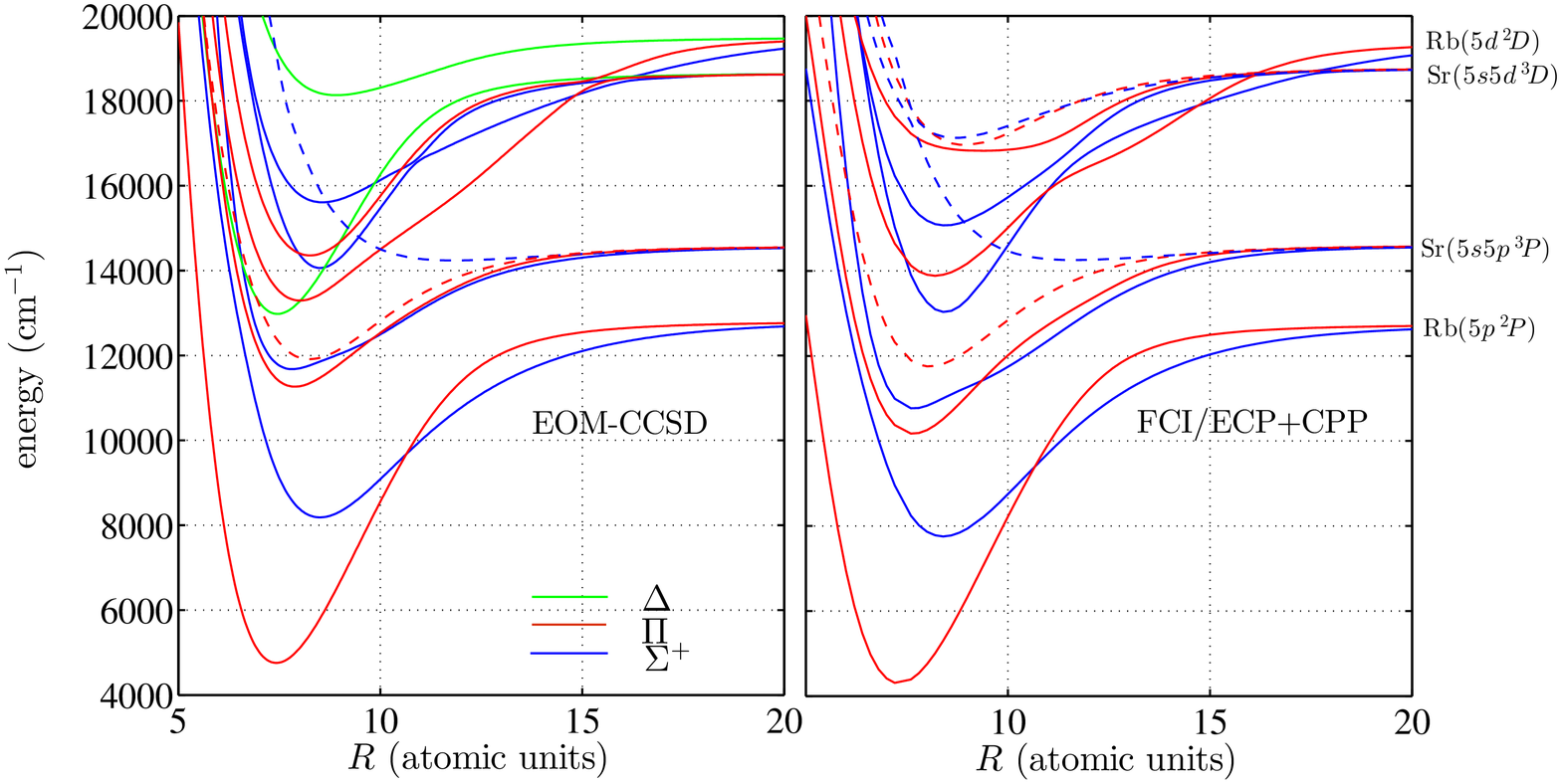}
\label{comparison_hund_a}
\end{figure*}

The excited state PECs calculated with both methods presented in Section \ref{methodssec} are displayed in Figure \ref{comparison_hund_a}, while in Table \ref{specconst} we report the main spectroscopic parameters of the Hund's case (a) PECs  correlated to Rb($^2S$)+Sr($^3P_{0,1,2}$) and Rb($^2P$)+Sr($^1S$).
%
%

\subsection{The Rb($5p\,^2P$)--Sr($^1S$) interaction}
 The two RbSr Hund's (a) case states correlated to this limit are denoted as $2^2\Sigma^+$ and $1^2\Pi$ . By construction the FCI/ECP+CPP method involves the exact asymptotic energy of 12737~cm$^{-1}$ (deduced from the position of $P$-state multiplet and Land\'e rule) while the EOM-CCSD method yields 12793~cm$^{-1}$ in good agreement (better than 0.5\%) with the former value. The overall agreement for the main spectroscopic quantities between FCI/ECP+CPP and EOM-CCSD PECs is satisfactory (see Table \ref{specconst}). Just like for the ground state PEC, the FCI/ECP+CPP method gives equilibrium distances shorter by about 0.1$a_0$ compared to the EOM-CCSD ones. This is due to the modeling of the short-range core-core repulsion that can be assumed to be identical to the one used for the ground state. This feature is also partly responsible for the deeper well depth (by about 5-7\%) and the smaller harmonic constant (by about 5\%) obtained with the FCI/ECP+CPP method for both $\Sigma$ and $\Pi$ states compared to the EOM-CCSD results. Note that both methods place the crossing between the $^2\Sigma^+$ and $^1\Pi$ states at almost the same distance: 10.67~$a_0$ in case of the FCI/ECP+CPP and 10.65~$a_0$ for the EOM-CCSD method. 

\begin{figure}
\caption{ Permanent electric dipole moments for 
(a) the $2^2\Sigma^+$ and $1^2\Pi$ states correlated to Rb($5p\,^2P$)--Sr($5s\,^1\!S$), 
(b) the $3^2\Sigma^+$ and $2^2\Pi$ states correlated to Rb($5s\,^2\!S$)+Sr($5s5p\,^3\!P$),
(c) the $1^4\Sigma^+$ and $1^4\Pi$ states correlated to Rb($5s\,^2\!S$)+Sr($5s5p\,^3\!P$), calculated with FCI/ECP+CPP and EOM-CCSD approaches. Solid lines: $^{2,4}\Sigma^+$ symmetry; dashed lines: $^{2,4}\Pi$ symmetry.}
\includegraphics[width=1.1\linewidth]{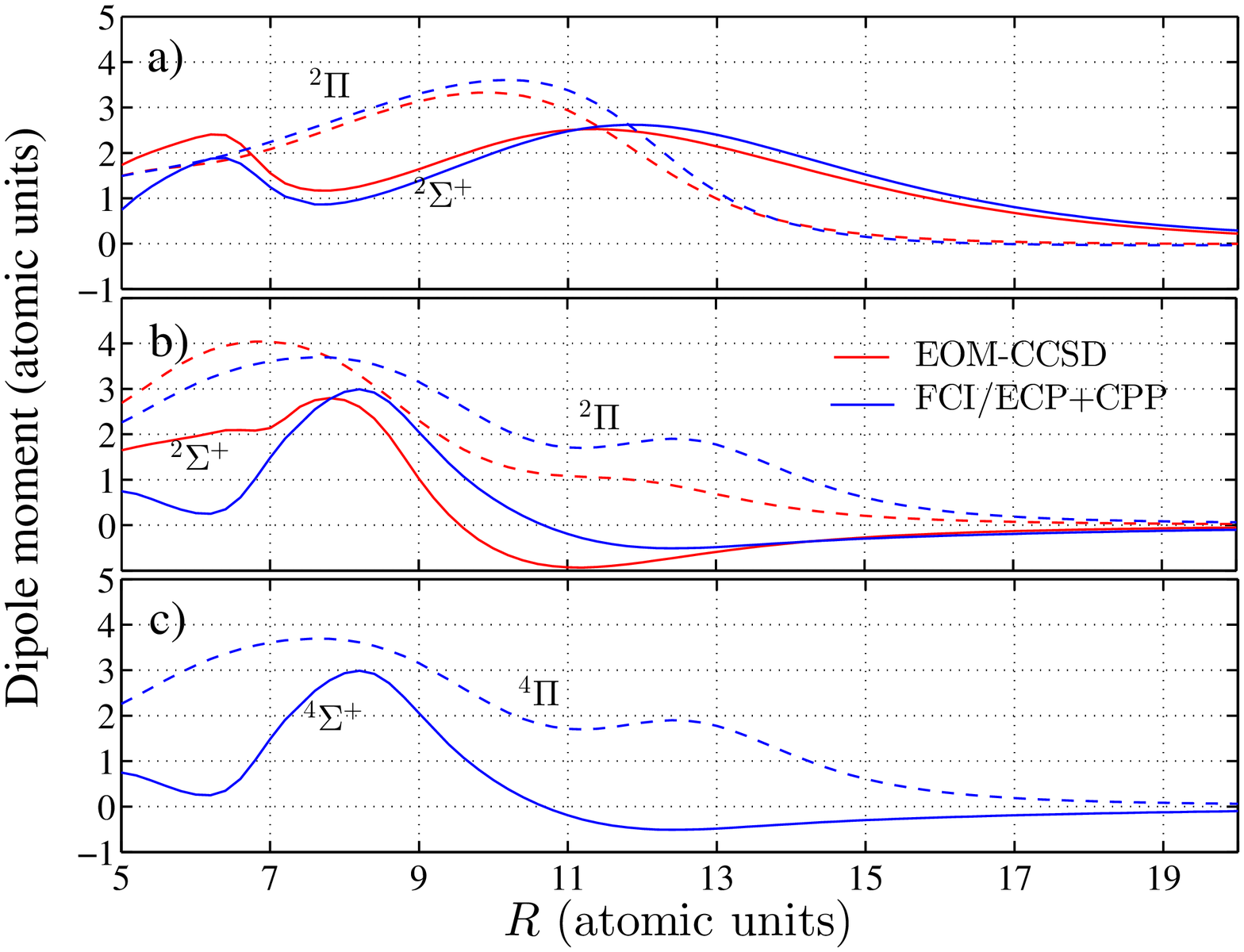}
\label{RbSr_exc_pdm}
\end{figure}

\begin{figure}
\caption{ Transition dipole moments from the RbSr ground state $X^2\Sigma^+$ towards (a) the $2^2\Sigma^+$ state (blue lines) and the $1^2\Pi$ (red lines) state, (b) the $3^2\Sigma^+$ state (blue lines) and the $2^2\Pi$ (red lines) state, obtained with EOM-CC (full lines), and with FCI/ECP+CPP (dashed lines).} 
\includegraphics[width=\linewidth]{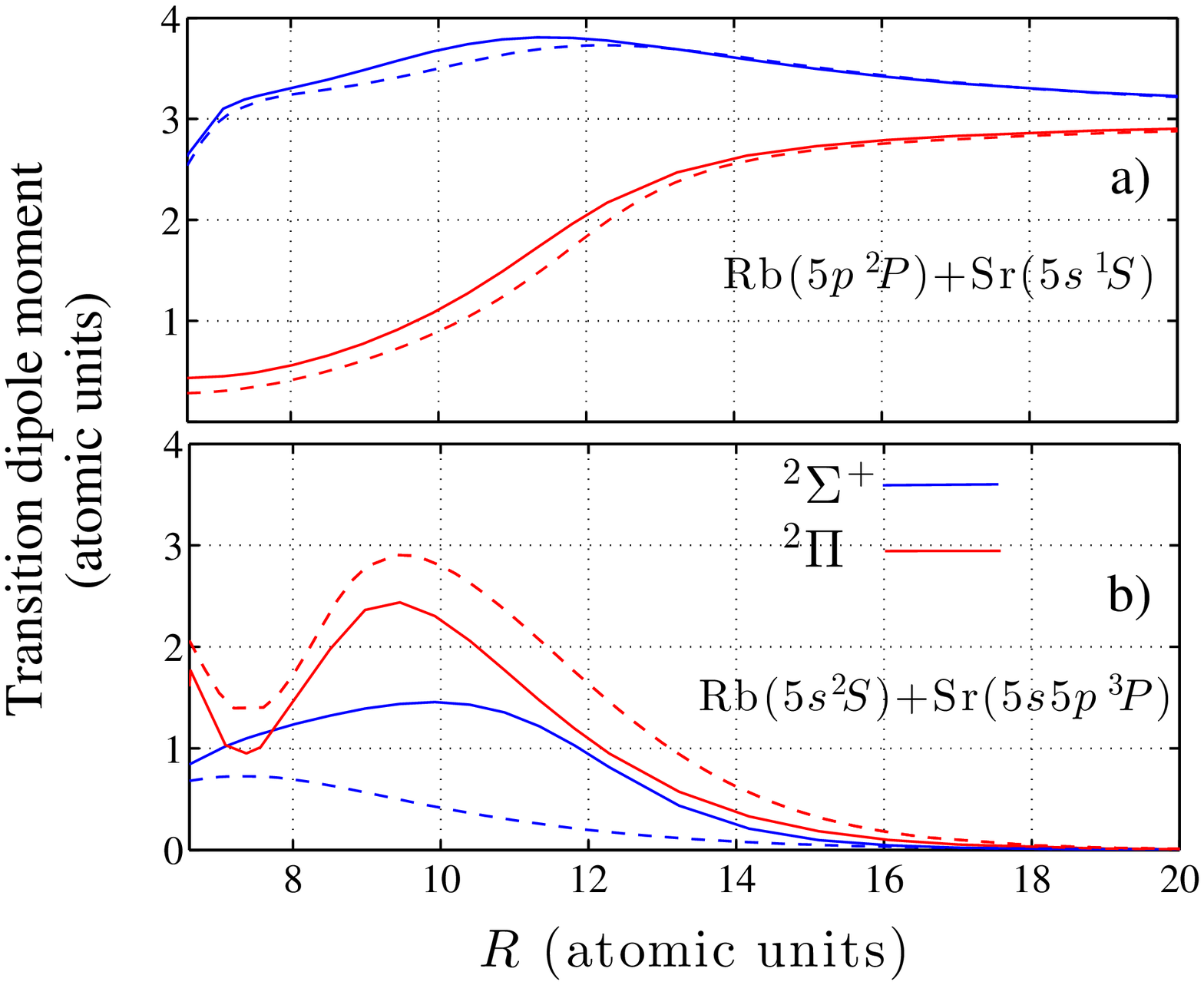}
\label{RbSr_exc_tdm}
\end{figure}

A very good agreement is found between the two methods on the PEDM of both the $2^2\Sigma^+$ and $1^2\Pi$ states (Fig. \ref{RbSr_exc_pdm}a) demonstrating again that both methods indeed describe the same electronic wave functions. The positions of maximum values of the PEDMs agree within 0.1 $a_0$, whereas their (large) magnitudes at this point agree to better than 5\%. The existence of two maxima in the $2^2\Sigma^+$ PEDM is probably related to a sudden change of chemical character of RbSr  molecule near the repulsive wall into an ion-pair state. As can be expected from the previous results on PEDMs, the agreement on the TDM functions for these states (Fig. \ref{RbSr_exc_tdm}a) is also excellent between the two approaches, as they involve ground and excited state wave functions which are represented in almost identical ways. Note that the asymptotic limit of these TDMs (3.013~a.u.) calculated at very long range agrees very well with the experimental atomic value $^2S \to ^2P_{\frac{1}{2}}$ transition (2.99 a.u.). At short distances the $X^2\Sigma^+ \to 1^2\Pi$ transition is clearly favored with respect to the $X^2\Sigma \to 2^2\Sigma$ transition.
\begin{table*}[htbp]
\centering
\caption{Main spectroscopic parameters of the  lowest excited states of RbSr Hund's case (a) potential energy curves}
\begin{ruledtabular}
\begin{tabular}{l|ccc|ccc|c}
          &    \multicolumn{3}{c}{ FCI/ECP+CPP} &     \multicolumn{3}{c}{EOM-CC} & asymptote \\
  State          &$R_e(a_0)$&$D_e$ (cm$^{-1}$)&$\omega_e$ (cm$^{-1}$)& $R_e$ ($a_0$)&$D_e$ (cm$^{-1}$)&$\omega_e$ (cm$^{-1}$)&   \\ \hline
 $1^{2}\Sigma^+$ & 8.69   &  1073.3   &  38.98  &  8.82  &       1040.5    &      38.09  & Rb($5s^2S$)+Sr($^1S$)   \\
 $2^{2}\Sigma^+$ & 8.40   &  4982.9   &  58.37  & 8.51   &       4609.6    &      60.20  & Rb($5p\,^2P$)+Sr($^1S$) \\
 $3^{2}\Sigma^+$ & 7.67   &  3828.0   &  65.26  & 7.81   &       2892.4    &      62.48  & Rb($5s^2S$)+Sr($5s5p ^3P$) \\
 $1^{2}\Pi$      & 7.31   &  8439.8   &  79.50  & 7.42   &       8038.6    &      83.19  & Rb($5p\,^2P$)+Sr($^1S$) \\
 $2^{2}\Pi$      & 7.65   &  4421.2   &  67.60  & 7.88   &       3303.5    &      63.37  & Rb($5s^2S$)+Sr($5s5p^3P$) \\ 
\hline
 $1^{4}\Sigma^+$ & 11.63  &  336.3   &  15.42  &  11.81   &        329.2    &      15.03$^\star$  &  Rb($5s$)+Sr($5s5p$) \\
 $1^{4}\Pi$ &  8.06  & 2838.1     &  56.98    &   8.24   &       2655.7    &      54.95$^\star$   &  Rb($5s$)+Sr($5s5p$)  \\
\end{tabular}
\end{ruledtabular}
\label{specconst}
\end{table*}

\subsection{The Rb($5s\,^2S$)--Sr($5s5p\,^3P$) interaction}
Four Hund's case (a) molecular states are correlated to this asymptote, which are denoted as $3\,^2\Sigma^+$, $1\,^4\Sigma^+$, $2\,^2\Pi$, and $1\,^4\Pi$.
All four states can easily be calculated with the FCI/ECP+CPP approach, while the computing codes for open-shell EOM-CCSD method for spin-changing states are not available. As mentioned in Section \ref{methodssec}, the $1\,^4\Sigma^+$ and $1\,^4\Pi$ quartet states correlating with Rb($^2S$)+Sr($^3P$) are the lowest ones for given spatial and spin symmetries, and are dominated by a single electronic configuration so that their PECs can be obtained with the RCCSD(T) method. 

The asymptotic limit of obtained excitation energies for these states are correctly reproduced to better than 1\% by our calculations when compared to the experimental value 14705~cm$^{-1}$  deduced from atomic data through the Land\'e rule for the $^3P$ strontium multiplet. The FCI/ECP+CPP method yields  14615~cm$^{-1}$ \cite{Guerout:2010} by construction for both the quartet and doublet states. The EOM-CCSD value for doublet states is 14567.8~cm$^{-1}$. 

Being the lowest states of their symmetry, the main spectroscopic constants for the $1\,^4\Sigma^+$ and $1\,^4\Pi$ PECs show a good agreement between the two methods, similar to the one obtained for the ground state PEC (Table \ref{specconst}). In contrast we immediately see from the Table and from Figure \ref{comparison_hund_a} that larger discrepancies are found between the two methods for doublet states. The equilibrium distance is now shorter by about 0.2$a_0$ in the FCI/ECP+CPP method, which could partly be accounted for the larger contribution of the core-core repulsion around 7.6$a_0$ than in the lowest states with a higher $R_e$. While consistent in magnitude, the harmonic constants differ by about 4~cm$^{-1}$ between the two methods, which could be reduced if we account for the short-range core-core term. The largest discrepancy concerns the well depth of the $3\,^2\Sigma^+$ and $2\,^2\Pi$ states which is deeper by about 1000~cm$^{-1}$ in the FCI/ECP+CPP results. The PEDM functions of Figure \ref{RbSr_exc_pdm}b reveal that while having similar trends, the details of the electronic wave functions induce significant differences in the oscillating patterns, \textit{i.e.} in the relative weights of the configurations. Obviously this feature transfers into the TDM functions of Figure \ref{RbSr_exc_tdm}b, in particular for the $X\,^2\Sigma^+ \to 3\,^2\Sigma^+$ transition where the magnitude of the TDM is weaker by about a factor of 2 in the FCI/ECP+CPP results compared to the EOM-CC ones, probably related to the different position of the node visible in the $3\,^2\!\Sigma^+$ PEDM. Nevertheless these TDMs deserve more attention. Asymptotically such transitions are in principle forbidden due to atomic spin-flip selection rule. Actually the $^1S \to ^3\!P_1$ excitation is allowed by electric dipole transition  due to second-order spin-orbit mixing with the higher $^1P_1$ state, while the excitation towards the $^3P_{0,2}$ states is strongly forbidden. In the molecular region, the disallowed transitions from ground state becomes allowed due to the Pauli exchange interaction which can be measured by the magnitude of the exchange energy \cite{Jeziorski:1976}).  The quite sudden increase of the TDM around 15$a_0$ reflects the exponential variation of this exchange energy when the valence-overlap region is reached.

Further studies with EOM-CC method which includes triply excited clusters and all-electron relativistic studies could probably help to validate one of these results.  Note that in the FCI/ECP+CPP approach the $1\,^4\Pi$ state is well separated in energy from other states to which it is coupled by spin-orbit interaction (see next Section), while it crosses the doublet states in the EOM-CC method. If the former result would be confirmed, this means that strongly polar metastable RbSr molecules could be created and used for further manipulation in the experiments.

%
\subsection{Higher excited states}
\label{higherstates}
Higher excited states of RbSr cannot be disregarded from the present analysis. An inspection of Figure \ref{comparison_hund_a} shows that the $3^2\Pi$ and $4^2\Sigma^+$ PECs correlated to the Rb($5s\,^2S$)+Sr($5s4d\,^3D$) dissociation limit  are submerged below the Rb($5s\,^2S$)+Sr($5s5p\,^3P$) asymptote by several hundreds of cm$^{-1}$ in both methods. The bottom of the well of the $1\,^2\Delta$ state (computed with the EOM-CCSD approach) is submerged by more than 1000~cm$^{-1}$ below that asymptote. 
Moreover, due to the proximity of their asymptotes, the excited states which correlate with Rb($4d\,^2D$)+Sr($5s^2\,^1S$), Rb($5s\,^2S$)+Sr($5s4d\,^3D$), and Rb($5s\,^2S$)+Sr($5s4d\,^1D$) strongly mix together and exhibit numerous avoided crossings which are consistently predicted by both methods: around 16$a_0$ for the $^2\Pi$ states, and around 16$a_0$ (EOM-CCSD) or 18$a_0$ (FCI/ECP+CPP) for the $^2\Sigma^+$ states. Both approaches also predict the presence of short-range avoided crossings of $\Pi$ and $\Sigma^+$ states but with more more pronounced differences
in positions. Finally, we observe  the large difference in the potential wells depths for $\Pi$ and $\Sigma^+$ states correlated to Rb($4d\,^2\!D$)+Sr($5s^2\,^1\!S$) and Rb($5s\,^2\!S$)+Sr($5s4d\,^3\!D$).
\subsection{Long-range behavior close to the Rb($5p\,^2P$)+Sr($5s^2\,^1S$) and Rb($5s\,^2S$)+Sr($5s5p\,^3P$) asymptotes}


We have obtained the $C_6$ values for the excited states by fitting the calculated potential energy curves at long range to $C_6 R^{-6}$ analytic form. This procedure has to be performed very carefully: backing out the van der Waals coefficients from the potential curves needs very high precision  potential curves for broad range of distances. As invoked in Section \ref{methodssec} due to the lack of high angular momentum functions in the basis set used in the FCI/ECP+CPP computations, we performed such fittings only for the CC methods. For the doublet states correlated to the Rb($5p\,^2\!P$)+Sr($5s^2\,^1\!S$) limit we obtained $C_6(2\,^2\Sigma^+)=23324 E_h a_0^6$ and $C_6(1\,^2\Pi)=8436 E_h a_0^6$ compared to the values of Ref.\cite{Jiang:2013}, 17530 $E_h a_0^6$ and 8331$E_h a_0^6$, respectively. Despite the nice agreement obtained for the latter value,  error bars for these values can be large and unpredictable as these values were obtained by fitting to the shape of  EOM-CCSD  potential curve added to the ground state interaction energy of RbSr. In contrast, the $C_6$ values for the excited states correlating with Rb($5s\,^2S$)+Sr($5s5p\,^3P$) can be extracted with much better accuracy, since for the quartet states potential energy curves  are  obtained in  direct  way and not as a sum of interaction energy + EOM-CC excitation. We obtained $C_6(2\,^2\Sigma^+)=5265 E_h a_0^6$ and $C_6(1\,^2\Pi)=4654 E_h a_0^6$ which are in satisfactory agreement with the values  5735$E_h a_0^6$ and 5000$E_h a_0^6$ of Ref.\cite{Jiang:2013}.

\subsection{Relativistic picture of the lowest excited states of RbSr.}  


 Spin-orbit (SO) splittings are quite large for the lowest excited states of both atoms: 237.6~cm$^{-1}$ for Rb($5p\,^2\!P$) and 581.1~cm$^{-1}$ for Sr($5s5p\,^3\!P$). Therefore they must be taken in account in any accurate representation of the RbSr excited states for the purpose of modelling experimental results. It is well-known that due to configuration mixing the SO couplings vary with the internuclear distance and can be reduced or enhanced typically by 30-50\% compared to the atomic values. Examples can be found for instance in spectroscopic studies of RbCs \cite{Docenko:2010}, KCs \cite{Kruzins:2010}, or in quantum chemistry studies of Sr$_2$ \cite{Tomza:2011}, SrYb \cite{Tomza:2012} or  Rb$_2$ \cite{Skomorowski:2012:jcp}. It is beyond the goal of this paper to compute the $R$-dependence of the SO coupling in RbSr. Instead we present an approximate model where the atomic SO is used as a perturbation to the Hund's case (a) states, in order to deliver a preliminary picture of the relevant PECs. Due to the large energy separation of the Rb($5p\,^2\!P$)+Sr($^1\!S$) and Rb($5s\,^2\!S$)+Sr($5s5p\,^3\!P$) asymptotes, the corresponding manifold of PECs can safely be considered as isolated from each other. We will also ignore the higher excited states discussed above which are submerged below the Rb($5s\,^2\!S$)+Sr($5s5p\,^3\!P$) asymptote.

We follow the usual spectroscopic convention and use symbols  $\Lambda$, $\Sigma$ for the projection onto the molecular axis of the electronic quantities namely the orbital angular momentum and the spin, respectively, and $\Omega= |\Lambda+\Sigma|$. The atomic SO constants are $A_{\rm Rb}=\Delta E_{\rm fs}({\rm Rb}(5p\,^2\!P))/3=79.2$~cm$^{-1}$ and $A_{\rm Sr}=\Delta E_{\rm fs}({\rm Sr}(5s5p\,^3\!P))/3=193.7$~cm$^{-1}$. We use the fact that the matrix elements of spin-orbit Hamiltonian $H_{\rm SO} = A \mathbf L \cdot\mathbf S$ in the basis $|S L \Sigma \Lambda> $ can be expressed in the {\em asymptotic} basis set of atomic angular momenta using Clebsch-Gordon   coefficients:
\begin{widetext}
\begin{equation}
|S L \Sigma \Lambda>  = \sum_{\Sigma_{\rm Rb},\Sigma_{\rm Sr} }  < S_{\rm Rb}  \Sigma_{\rm Rb} S_{\rm Sr}   \Sigma_{\rm Sr}  | S \Sigma >   |S_{\rm Rb}  \Sigma_{\rm Rb}> |S_{\rm Sr}  L \Sigma_{\rm Sr}  \Lambda >,
\label{S1S2_cpl}
\end{equation}
which more specifically reduces for doublet states to 
\begin{equation}
|\frac{1}{2} L  \pm \frac{1}{2} \Lambda > =
\pm  \sqrt{\frac{1}{3}} |\frac{1}{2}  \frac{1}{2} >_{\rm Rb}  |1 L 0 \Lambda >_{\rm Sr} 
 \mp\sqrt{\frac{2}{3}}  |\frac{1}{2}  -\frac{1}{2} >_{\rm Rb}  |1 L 1 \Lambda >_{\rm Sr},
\end{equation}
while for the quartet states it reads
\begin{eqnarray}
|\frac{3}{2} L  \pm \frac{1}{2} \Lambda > &=&
 \sqrt{\frac{1}{3}} |\frac{1}{2}  \frac{1}{2} >_{\rm Rb}  |1 L 0 \Lambda >_{\rm Sr} 
 +\sqrt{\frac{2}{3}}  |\frac{1}{2}  -\frac{1}{2} >_{\rm Rb}  |1 L 1 \Lambda >_{\rm Sr}, \\
|\frac{3}{2} L  \pm \frac{3}{2} \Lambda >&=&
  |\frac{1}{2}  \pm \frac{1}{2} >_{\rm Rb}  |1 L \pm1 \Lambda >_{\rm Sr}. 
\end{eqnarray}
\end{widetext}
The interaction of  rubidium atom in the $^2\!P$ state  with Sr ground state atom splits the degeneracy of the $^2P$ state into $^2\Sigma$ and $^2\Pi$ state. The total angular momentum projection $|\Omega|$ can then take the values $\frac{1}{2}$ and $\frac{3}{2}$. A unique state $|\Omega|=\frac{3}{2}$ originates from $^2\Pi(\Sigma=\pm\frac{1}{2},\pm \Lambda=1)$ state, while two states with  $|\Omega|=\frac{1}{2}$ states originate from mixing of the $^2\Pi(\Sigma=\pm\frac{1}{2}, \Lambda=\mp 1)$ and $^2\Sigma^+(\Sigma=\pm\frac{1}{2}, \Lambda=0)$.  The Hamiltonian for the $|\Omega|=\frac{3}{2}$  state is trivially reduced to one element only, which can be written as $H(|\Omega|=\frac{3}{2})=V(^2\!\Pi) + 2 A_{\rm Rb} $ and asymptotically corresponds to $j=\frac{3}{2}$ state of the Rb atom. For the $|\Omega|=\frac{1}{2}$ state the Hamiltonian can be written as
\begin{equation}
H(|\Omega|=\frac{1}{2}) = \left(
\begin{array}{cc}
 V(2^2\Sigma) &  \sqrt{2} A_{\rm Rb} \\
  \sqrt{2} A_{\rm Rb}  &  V(1^2\Pi)  +  A_{\rm Rb}
\end{array}\right).
\end{equation}
Two eigenvalues of this matrix asymptotically correspond to both $j=\frac{1}{2}$, $\frac{3}{2}$, state of excited Rb($5p$) atom.

For the interaction of $^3\!P$ state of Sr and $^2\!S$ of Rb the situation is somewhat more complicated. As we have mentioned in previous sections, the resulting dimer states for the Rb($5s\,^2\!S$)--Sr($5s5p\,^3\!P$) interaction in the Hund's case (a) are $^{2,4}\Sigma^+$ and $^{2,4}\Pi$. The possible quantum numbers for spin-orbit coupled states for that case are  $|\Omega|=\frac{1}{2}$, $\frac{3}{2}$ and $\frac{5}{2}$. The maximal value of $|\Omega|$ corresponds trivially to single state, namely $H(|\Omega|=\frac{5}{2})=V(1^4\!\Pi) + A_{\rm Sr}$ asymptotically corresponding to the metastable state of Sr atom $^3\!P_2$. The $|\Omega|=\frac{3}{2}$ states can be obtained by coupling of  three states: $^4\Sigma^+(\Sigma=\pm\frac{3}{2}, \Lambda=0)$, $^2\Pi(\Sigma=\pm\frac{1}{2}, \Lambda=\pm 1)$ and $^4\Pi(\Sigma=\pm\frac{1}{2}, \Lambda=\pm 1)$. The corresponding Hamiltonian from which we can obtain the Hund~(c) case representation reads:
\begin{widetext}
\begin{eqnarray}
H(|\Omega|=\frac{3}{2}) = \left(
\begin{array}{ccc}
 V(^2\Pi) +\frac{2}{3} A_{\rm Sr}  &   \sqrt{\frac{1}{3}} A_{\rm Sr}   &  -\frac{\sqrt{2}}{3} A_{\rm Sr}  \\
\sqrt{\frac{1}{3}} A_{\rm Sr}         &  V(^4\Sigma)                         &   \sqrt{\frac{2}{3}} A_{\rm Sr}  \\
 -\frac{\sqrt{2}}{3} A_{\rm Sr}       &   \sqrt{\frac{2}{3}} A_{\rm Sr}   &    V(^4\Pi) +\frac{1}{3} A_{\rm Sr}   \\
\end{array}
\right).
\end{eqnarray}
\end{widetext}
Two of these states asymptotically correspond to the $^3\!P_2$ Sr state, and one to the $^3\!P_1$ Sr state. Finally, for the $|\Omega|=\frac{1}{2}$ we have five states involved:   $^2\Sigma^+(\Sigma=\pm\frac{1}{2}, \Lambda=0)$,  $^2\Pi(\Sigma=\mp\frac{1}{2}, \Lambda=\pm 1 )$, $^4\Sigma^+(\Sigma=\pm\frac{1}{2}, \Lambda=0)$,  $^4\Pi(\Sigma=\mp\frac{1}{2}, \Lambda=\pm 1)$, $^4\Pi(\Sigma=\pm\frac{3}{2}, \Lambda=\mp 1)$. The  Hamiltonian which describes the coupled $|\Omega|=\frac{1}{2}$ states has the following form:
\begin{widetext}
\begin{eqnarray}
H(|\Omega|=\frac{1}{2}) = \left(
\begin{array}{ccccc}
 V(^2\Sigma^+)                           &    \sqrt{\frac{8}{9}}  A_{\rm Sr}         &   	0           			        	 &	-\frac{1}{3}A_{\rm Sr}                          & 	\sqrt{\frac{1}{3}}A_{\rm Sr}	 	 \\
   \sqrt{\frac{8}{9}}  A_{\rm Sr}       & V(^2\Pi) -\frac{2}{3} A_{\rm Sr}        &     \frac{1}{3}A_{\rm Sr}       	&    	-\frac{\sqrt{2}}{3}A_{\rm Sr}                  &   				0			  	 \\
   	0	                            	    & 	  \frac{1}{3}A_{\rm Sr} 	      &   V(^4\Sigma^+)     			 &     \sqrt{\frac{8}{9}}  A_{\rm Sr}	               & 	  \sqrt{\frac{2}{3}}  A_{\rm Sr} 	 \\
-\frac{1}{3}A_{\rm Sr}       		    & 	-\frac{\sqrt{2}}{3}A_{\rm Sr} 	    & \sqrt{\frac{8}{9}}  A_{\rm Sr}    &  V(^4\Pi)   -\frac{1}{3}  A_{\rm Sr}  & 		0						  \\
  \sqrt{\frac{1}{3}}A_{\rm Sr}    	    & 	0	  					    & \sqrt{\frac{2}{3}}  A_{\rm Sr}    &   		0	                                     & V(^4\Pi) -  A_{\rm Sr}		  \\
\end{array}
\right).
\end{eqnarray}
\end{widetext}
The eigenstates of the Hamiltonian  for $\Omega=\frac{1}{2}$ correspond to all components of the $^3P$ asymptote of the excited Sr atom: the lowest eigenvalue represents the interaction of Rb atom with $^3P_0$ state of Sr, two states correspond to the interaction with $^3P_1$, and two with $^3P_2$. 
\begin{figure*}[htp]
\caption{
EOM-CCSD potential energy curves of RbSr excited states including atomic spin-orbit interaction as explained in the text. (a) EOM-CC approach, (b) FCI/ECP+CPP approach.}
\includegraphics[width=\linewidth]{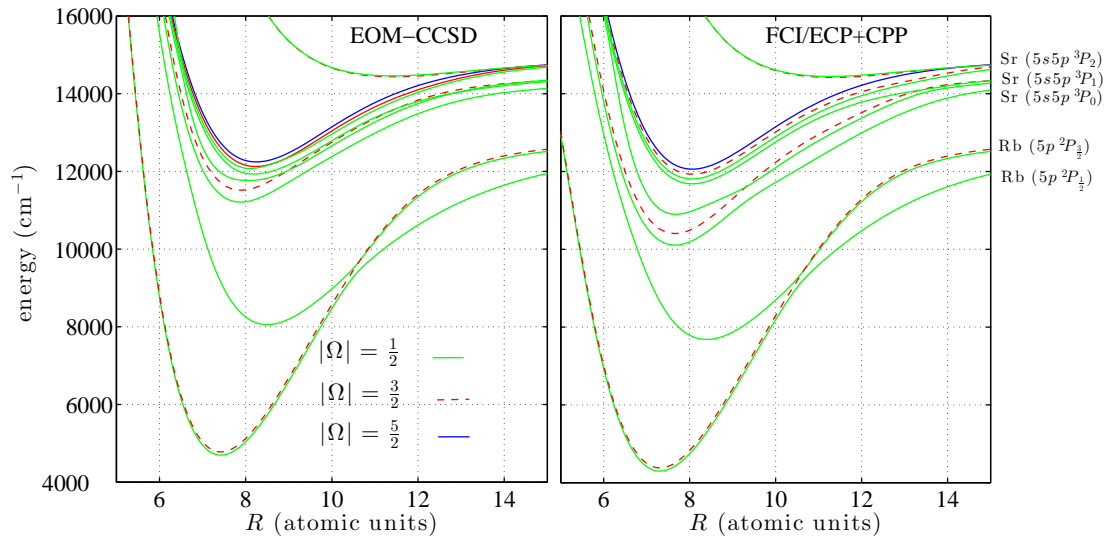}
\label{hundc_vs_hunda}
\end{figure*}
The resulting spin-orbit coupled PECs are shown in Figure \ref{hundc_vs_hunda}. The EOM-CC and FCI/ECP+CPP approaches give a very consistent potential energy curves, except some of the states which correlate with the Rb($5s\,^2S$)+Sr($5s5p\,^3P_{0,1,2}$) asymptote, which originate from doublet $3^2\Sigma^+$ and $2^2\Pi$ states which are about 20\% deeper in case of the FCI/ECP+CPP method.	

For the Rb($5p\,^2P_{1/2,3/2}$)--Sr($^1S$) manifold the $|\Omega|=\frac{1}{2}$ curves exhibit the avoided crossing in place where the $1^2\Pi$ and $2^2\Sigma^+$ states cross. Since the $1^2\Pi$ and $2^2\Sigma^+$ states are separated in energy by a much larger amount than the SO constant $A_{\rm Rb}$, they preserve their Hund (a) case character over most of the internuclear distances in the chemical range. However the corresponding bound levels may well be strongly coupled as it is the case for instance in the heavy alkali-metal dimers like Rb$_2$ \cite{Amiot:1999}. For the states that correlate with the Rb($5s\,^2S$)+Sr($5s5p\,^3P_{0,1,2}$) asymptote the character of Hund's case (c) states is  drastically changed, since the $2^2\Pi$ and $3^2\Sigma^+$, and also $1^4\Pi$ states are much closer in energy. Therefore these states are strong mixtures of doublet and quartet states. It is also clear from Figure \ref{hundc_vs_hunda} that among the states which correlate with Rb($5s\,^2S$)+Sr($5s5p\,^3P_{2}$) there are states with  $|\Omega|=\frac{1}{2}$  and $|\Omega|=\frac{3}{2}$  of very strong $^4\Sigma^+$ character. Also, in view of such strong mixing, it is clear that the state $|\Omega|=\frac{1}{2}$ which correlates with Sr atomic clock line will have nonzero transition dipole moment at finite distances. Hence, the vibrational states supported by such state might be accessible with dipole transitions.

\section{Conclusions and outlook}
In this work we have explored the  ground and excited states of the RbSr molecule, which is a good candidate  paramagnetic, polar molecule and subject to intense experimental study.
A primary goal of this paper was to provide the first calculation of  potential energy curves for the RbSr curves and dipole moment matrix elements and comparison between two {\em ab-initio} methods:
 FCI method with the use of ECP and CPP, and CC theory based methods used with small effective core potential. 

It is usually difficult to provide error bound for the {\em  ab-initio} calculations, unless we deal with small, few-electron system \cite{Przybytek:2005}, for which it is possible to study the convergence pattern not only for systematically increased gaussian basis set but also for number of excitations introduced to the electronic wavefunction. Hence, for such complicated system as RbSr molecule application of two different methods provides a better starting point for further modelling of potential curves with help 
of high-resolution spectrosopy experiments. The discrepancies in calculations of potential energy curves between the methods used in this paper are very small for the ground state, for the states correlating with Rb asymptotes and for the quartet states correlating with the strontium asymptotes. A bit larger discrepancies have been obtained for the doublet states of Rb+Sr($5p$) systems, although the equilibrium distances and harmonic constants are consistent.  For the higher excited states the agreement is moderately good. 

Using both methods we have found that a very good agreement of  the values of permanent dipole moments of the ground state RbSr system as well as doublet Rb($5p$)+Sr and Rb+Sr($5p$) states: interestingly enough, the permanent dipole moments of the excited states are very large.

 Finally, we have obtained the transition dipole moments for the excitations from the ground electronic state to Rb($5p$)+Sr and Rb+Sr($5p$) states. Again, there is a good agreement between the two approaches used in this paper. Interestingly enough we have found that there are non-zero transition dipole moments from the ground state to doublet Rb+Sr($5p$). That means, the possibility of driving 
for the dipole transitions to the vibrational states supported by  the electronic states that correlate with strongly forbidden $^3P_J$ lines of the strontium atom. 


\section{Acknowledgements}
This work was supported in part by the National Science Foundation (Grant No.~NSF PHY11-25915). RG acknowledges support from the \textit{Institut Francilien de Recherches sur les Atomes Froids}. PSZ is grateful for the support of the Foundation for Polish Science Homing Plus Programme (2011 3/14) co-financed by the European Regional Development Fund. The authors are also grateful to Florian Schreck and Benjamin Pasquiou for  sparking their interest in the RbSr system.
	
%

\end{document}